\documentstyle[aps,preprint,axodraw]{revtex}
\newcommand{\beq}{\begin{equation}}
\newcommand{\eeq}{\end{equation}}
\newcommand{\beqa}{\begin{eqnarray}}
\newcommand{\eeqa}{\end{eqnarray}}
\newcommand{\beqan}{\begin{eqnarray*}}
\newcommand{\eeqan}{\end{eqnarray*}}

\begin{document}

\title{\rightline{\small SINP/TNP/99-13}
\rightline{\small SBNC/99/04-01}
{\bf Fermion Decoupling and the Axial Anomaly on the Lattice}}

\author{{\bf H. Banerjee$^a$ and Asit K. De$^b$} \\
$^a${\sl S.N.Bose National Centre for Basic Sciences,
JD Block, Salt Lake, Calcutta 700 091, India\\
e-mail:banerjee@boson.bose.res.in\\} 
$^b${\sl Theory Group, Saha Institute of Nuclear Physics,
1/AF Salt Lake, Calcutta 700 064, India\\
e-mail:de@tnp.saha.ernet.in}}

\date{March,1999}


\maketitle

%


\begin{abstract}
By an explicit calculation of the continuum limit of the triangle graph 
amplitude in lattice QED
we show that in the axial Ward identity the Adler-Bell-Jackiw (ABJ)
anomaly exactly cancels the pseudoscalar density term
$2im\langle\overline{\psi}_x\gamma_5\psi_x\rangle_{a=0}$ in the limit of
infinite fermion mass $m$. The result, a reflection of decoupling of the heavy
fermion, provides a convenient framework for computing the flavor-singlet or
U(1) axial anomaly in non-abelian gauge theories on lattice. 
Our calculations on the lattice are performed using Wilson
fermions but the results are general.

\end{abstract}


\newpage
\noindent
{\bf Introduction.}
The most well-known model for lattice fermions, the Wilson model
\cite{Wilson77}, solves the problem of species doubling through an 
{\it irrelevant
term}, the Wilson term, which breaks chiral symmetry explicitly. Explicit
breaking of chiral symmetry, turns out to have a more
profound reason. According to popular perception \cite{KaSm81} it is also
necessary to generate, from the lattice regulated model,
the ABJ anomaly in perturbation theory in the continuum limit.
Indeed, the contribution of the Wilson term to the four-divergence of the
axial current is treated as the driving term for the ABJ anomaly
\cite{KaSm81,Kerler83}. 

To examine the role of the underlying lattice fermion model in generating
the ABJ anomaly a convenient and transparent starting point is the
condition, in this context,
for decoupling of the fermion in the large mass limit from the background
gauge field \cite{Apple},
\begin{equation}
\langle \Delta_\mu J_{\mu 5}(x)\rangle_{a=0}
=
2im \;\; \langle \overline{\psi}_x\gamma_5\psi_x\rangle_{a=0}
-\lim_{m\rightarrow \infty}\left[2im\langle\overline{\psi}_x\gamma_5\psi_x
\rangle_{a=0}\right],\label{decoupling}
\end{equation} 
where $a$ is the lattice constant. One recognises Eq.(\ref{decoupling}) as
the Adler condition \cite{Adler} which states that the triangle graph
amplitude should vanish in the limit as the mass of the loop fermion becomes
infinite. To establish that the decoupling condition is indeed equivalent to
the axial Ward identity one needs the supplementary relation
\begin{equation}
\lim_{m\rightarrow \infty}\left[2im\;\;\langle\overline{\psi}_x
\gamma_5\psi_x\rangle_{a=0}\right] = \frac{ig^2}{16\pi^2} 
\epsilon_{\mu\nu\lambda\rho}{\rm tr}F_{\mu\nu}F_{\lambda\rho}, \label{anomaly1}
\end{equation}
where $F_{\mu\nu}$ is the gauge field tensor. A point to note is that, it is
convenient to consider the continuum limit of the pseudoscalar density
$\langle\overline{\psi}_x\gamma_5\psi_x\rangle_a$ being of dimension three,
whereas the contribution of the irrelevant term in the axial Ward identity
on lattice is of dimension four.

Our derivation of Eqs.(\ref{decoupling}) and (\ref{anomaly1}) in lattice QED
demonstrates that as long as the underlying lattice fermion model removes
doubling completely and is gauge-invariant and local, the ABJ anomaly is
generated without reference to the specific form of the irrelevant term. In
non-abelian gauge theories on lattice Eq.(\ref{decoupling}) provides, as we
shall see, a simple recipe for deriving the U(1) or flavor-singlet axial
anomaly.

\newpage
\noindent
{\bf Decoupling in QED.}
The key to our analysis is the Rosenberg \cite{Rosen} tensor decomposition
of the amplitude of the triangle diagrams (i) and (ii) in continuum QED for
axial current $j_{\lambda 5}(x)$ to emit two photons with momenta and
polarisation ($p,\mu$) and ($k,\nu$).

\begin{equation}
T^{(i+ii)}_{\lambda\mu\nu}
=
\epsilon_{\lambda\mu\nu\alpha}  k_\alpha A(p,k,m) \;+\; 
\epsilon_{\lambda\nu\alpha\beta} p_\alpha k_\beta [p_\mu B(p,k,m)
\;+\; k_\mu C(p,k,m)] \;+\; [(k,\nu) \leftrightarrow (p,\mu)].
\label{rosen}
\end{equation}

Gauge invariance relates the Rosenberg form factors A to B and C
\begin{equation}
A(p,k,m)=p^2B(p,k,m)+p.k\; C(p,k,m).\label{gi}
\end{equation}

It is to be noted that the form factors $B$ and $C$ are of mass
dimension -2, and, therefore, must vanish as $m^{-2}$ for large fermion
mass. Gauge invariance, therefore, guarantees that  

\beqa
\lim_{m\rightarrow\infty} (p+k)_\lambda T^{(i+j)}_{\lambda\mu\nu}
& = &
- \epsilon_{\mu\nu\alpha\beta} p_\alpha k_\beta
\lim_{m\rightarrow\infty}[A(p,k,m)+A(k,p,m)]  \label{anomlim} \\
& = & 0 ,\label{proof}
\eeqa
which is the basis of Eq.(1). In the above, (\ref{proof}) follows from
(\ref{anomlim}) because of (\ref{gi}) and the asymptotic behavior of $B$ and
$C$.

On lattice, the decoupling condition (\ref{proof}) should be realised in the
continuum limit irrespective of the underlying model for fermion as long as
it is free from doublers and local. The form factors $B$ and $C$ which are
highly convergent amplitudes must coincide with their respective expressions
in the continuum in all {\em legitimate} lattice models. Residual model
dependence, if any, can appear only in the form factor $A$ because of
potential logarithmic divergence. This, however, is ruled out by the gauge
invariance constraint (\ref{gi}).

In lattice QED with Wilson fermions, 
the Feynman amplitudes corresponding to the two diagrams (i) and (ii) are :
\beqa
[T^{(i+ii)}_{\lambda\mu\nu}]_a & = & -g^2 
                     \int_{-\frac{\pi}{a}}^{\frac{\pi}{a}}
\frac{d^4l}{(2\pi)^4}       {\rm Tr}\Bigg[\gamma_\lambda\gamma_5 \cos
a(l+\frac{k-p}{2})_\lambda                      S(l-p) V_\mu(l-p,l) S(l)
                     V_\nu(l,l+k) S(l+k) \nonumber \\ 
& &
+ (k,\nu \leftrightarrow p,\mu)\Bigg],    
\label{ampli}
\eeqa
with the fermion
propagator $S(l)$ and the one-photon vertex $V_\mu(p,q)$ given by
\beqa
S(l) &=& \left[ \sum_\mu \gamma_\mu \frac{\sin al_\mu}{a} + 
          \frac{r}{a} \sum_\mu (1-\cos al_\mu)+m \right]^{-1}, \\
V_\mu &=& \gamma_\mu \cos \frac{a}{2}(p+q)_\mu + r \sin \frac{a}{2}(p+q)_\mu.
\eeqa
where $r$ is the Wilson parameter.

On lattice there are four additional diagrams with {\em irrelevant} vertices.
As will be evident from the following, they do not contribute in the continuum
limit.

The lattice amplitude (\ref{ampli}) is superficially linearly divergent.
However, the leading term, obtained by setting the
external momenta $p, \;k=0$ is odd in the loop momentum $l$ and, therefore,
vanishes due to symmetric integration. The amplitude, therfore, vanishes at 
least linearly in external momenta as indeed the Rosenberg decomposition 
suggests and, furthermore, the effective divergence is at most logarithmic.

Our strategy is to consider the derivative of (\ref{ampli}) with respect to the
fermion mass $m$ rather than the external momenta $p,\;k$ as is common practice
\beq
[R^{(i)}_{\lambda\mu\nu}]_a\equiv
\frac{d}{dm}[T^{(i)}_{\lambda\mu\nu}]_a . \label{R}
\eeq  
Lattice power counting gives a negative integer for the effective degree of
divergence of $[R^{(i)}_{\lambda\mu\nu}]_a$. 
One can, therefore, take, thanks to the Reisz theorem \cite{Reisz},
the continuuum limit of the integrands and evaluate the loop integrals in
the entire phase space $-\infty \le l_\mu \le \infty$ as in the continuum
Feynman amplitudes. In the continuum limit, amplitudes of only two 
diagrams (i) and (ii) survive and amplitudes with {\em irrelevant} vertices 
vanish. The amplitudes
$[R^{(i)}_{\lambda\mu\nu}]_{a=0}$ and $[R^{(ii)}_{\lambda\mu\nu}]_{a=0}$ 
are individually Bose-symmetric and hence gauge-invariant:
\beq
R^{(i+ii)}_{\lambda\mu\nu} = 2\;\;R^{(i)}_{\lambda\mu\nu} = 2\;\;
R^{(ii)}_{\lambda\mu\nu}.
\eeq  
which is an unexpected bonus.

The Rosenberg tensor decomposition for $[R^{(i+ii)}_{\lambda\mu\nu}]_{a=0}$ is given by 
\beqa
[R^{(i+ii)}_{\lambda\mu\nu}]_{a=0}
& = &
4g^2 m \int^\infty_{-\infty} \frac{d^4l}{(2\pi)^4}
\Bigg[{\rm Tr}(\gamma_5 \gamma_\lambda \gamma_\mu \gamma_\nu p\!\!/)
\left(\frac{1}{D}(1+\frac{k^2}{d_3})-\frac{1}{d_1d^2_3}\right) \nonumber \\
&  &+{\rm Tr}(\gamma_5\gamma_\lambda\gamma_\nu p\!\!/ k\!\!\!/)
\frac{2(l_\mu-p_\mu)}{Dd_1} + (p,\mu \leftrightarrow k,\nu)\Bigg]\label{Rosen}
\eeqa
where
\beq
D=d_1d_2d_3 \;\;\; {\rm and}\;\;\;
d_1\equiv (l-p)^2+m^2,\;\; d_2\equiv l^2+m^2,\;\; d_3\equiv (l+k)^2+m^2.
\eeq

The four-divergence of the amplitude for the axial vector current is to be obtained from
\beqa
[(p+k)_\lambda R^{(i+ii)}_{\lambda\mu\nu}]_{a=0}
& = & \frac{d}{dm}[(p+k)_\lambda T^{(i+ii)}_{\lambda\mu\nu}]_{a=0} \nonumber
\\
& = & -\frac{1}{\pi^2}\epsilon_{\mu\nu\alpha\beta}p_\alpha k_\beta
      \frac{d}{dm}\int_{0\le s+t\le 1}\frac{m^2}{c^2+m^2} ds\;dt,
\eeqa
with
\beq
c^2\equiv s(1-s)p^2 + t(1-t)k^2 +2st\;p.k.
\eeq

The Adler condition (1) determines the {\em constant of integration}
\beq
[(p+k)_\lambda T^{(i+ii)}_{\lambda\mu\nu}]_{a=0}
=
-\frac{1}{\pi^2}\epsilon_{\mu\nu\alpha\beta}p_\alpha k_\beta
\left[\int \frac{m^2}{c^2+m^2}ds\;dt -\frac{1}{2}\right] \label{div}.
\eeq 

The ABJ anomaly is identified as the $m=0$ limit of the right hand side of 
(\ref{div}):
\beq
{\rm ABJ\;\; anomaly} = \frac{1}{2\pi^2}\epsilon_{\mu\nu\alpha\beta}p_\alpha 
k_\beta \label{anomaly}
\eeq

\noindent
{\bf U(1) axial anomaly in non-abelian gauge theories.}
The representation motivated by the decoupling condition (1)
\beq
\lim_{m\rightarrow
\infty}\left[2im\langle\overline\psi_x\gamma_5\psi_x\rangle_{a=0}\right]
=
\lim_{m\rightarrow \infty}\left[2im\langle x|{\rm Tr}\gamma_5(D\!\!\!\!/
+W+m)^{-1)}|x\rangle_{a=0}\right] \label{PV}
\eeq
constitutes the starting point of our calculation of the axial anomaly in
non-Abelian theories, {\em e.g.}, lattice QCD. The Dirac operator 
$D\!\!\!\!/$ and the Wilson term $W$ are given by
\beqa
D_\lambda & \equiv & \frac{1}{2ia}\left(e^{ip_\lambda a} U_\lambda
                    -U_\lambda^\dagger e^{-ip_\lambda a}\right) \\ \nonumber
W &\equiv & \frac{r}{2a} \sum_\lambda \left(2-e^{ip\lambda a} U_\lambda
                     -U^\dagger_\lambda e^{-ip_\lambda a}\right)
\eeqa
where $U_\lambda\equiv exp(iagA_\lambda)$ is the link variable with
$A_\lambda \equiv t^a A^a_\lambda$ the gauge potetial and $t^a$ the
generators of $SU(N)$.

One recognises (\ref{PV}) as the analog of the contribution of the
Pauli-Villars fermion to the axial anomaly in continuum $SU(N)$ theory
\cite{Banerjee}. As in the continuum, our strategy is to develop the Green
function for lattice fermion in a perturbative series:
\beqa
(D\!\!\!\!/+W+m)^{-1} & = & (-D\!\!\!\!/+W+m)G \nonumber \\
{\rm with}\;\;\; G & = &\left(-D\!\!\!\!/^2+(W+m)^2+[D\!\!\!\!/,W]\right)^{-1} 
\nonumber \\
& = & G_0 -gG_0VG_0 +g^2 G_0VG_0VG_0+ ... \label{perturb}
\eeqa
where the {\em free} part 
\beq
G_0 = \left( \sum D^2_\mu +(W+m)^2\right)^{-1}
   = \left[\sum\frac{\sin^2ap_\mu}{a^2}+ \left(\frac{r}{a}\sum_\mu(1-\cos
ap_\mu)+m\right)^2\right]^{-1}
\eeq
is of Reisz degree $-2$ and has the expected continuum limit
\beq
[G_0]_{a=0} = (p^2+m^2)^{-1}
\eeq 

The potential $gV$ has three pieces
\beq
gV = gV_0 + gV_1 + gV_2
\eeq
of which the first piece $gV_0$ is independent of $\gamma$-matrices, has
Reisz degree $+1$ and non-vanishing continuum limit. The pieces $gV_1$ and
$gV_2$ contain $\gamma$-matrices and each has Reisz degree zero. The
continuum limit of $gV_1$ vanishes
\beq
(gV_1)_{a=0} = [D\!\!\!\!/,W]_{a=0} =0,
\eeq
whereas, 
\beq
(gV_2)_{a=0}= \frac{i}{2} \sigma_{\mu\nu} \left[D_\mu,D_\nu\right]_{a=0}
           = -\frac{i}{2} \sigma_{\mu\nu} F_{\mu\nu}
\eeq
where $F_{\mu\nu}$ is the field tensor in the continuum.

It is clear that the first two terms of the perturbative series
(\ref{perturb}) do not contribute to the axial anomaly (\ref{PV}) simply 
because they do not have enough $\gamma$-matrices to give non-vanishing
Dirac trace. Reisz power counting for the lattice Feynman amplitudes
corresponding to the second and higher order terms in (\ref{perturb}) all
give negative integers. One can now use the Reisz theorem and take the
continuum limit of the integrands in the lattice Feynman amplitudes
corresponding to all these terms. Since $gV_1$ vanishes in the continuum
limit it cannot contribute to the axial anomaly. The term which survive in
the large mass limit in the continuum is thus given by
\beqa
\lim_{m\rightarrow \infty}\left[2im\langle\overline\psi_x\gamma_5\psi_x
\rangle_{a=0}\right]
& = &
\lim_{m\rightarrow \infty} \left[ 2img^2\langle x|{\rm Tr} \gamma_5 G_0
V_2 G_0 V_2 G_0|x\rangle_{a=0}\right] \nonumber \\
& = &
\frac{ig^2}{16\pi^2}\epsilon_{\lambda\rho\mu\nu} {\rm tr} F_{\lambda\rho}(x)
F_{\mu\nu}(x) \label{final}
\eeqa
where `tr' now denotes trace over internal symmetry indices. Note that the
final result (\ref{final}) is local, all nonlocalities disappearing in the
large $m$ limit, as do all higher order terms in the perturbative series
(\ref{perturb}).

One of the authours (H.B.) is indebted to Kazuo Fujikawa for illuminating
correspondence.

\vskip 3cm

\begin{center}
\begin{picture}(500,200)(0,0)
\Vertex(80,175){2.5}
\Vertex(250,175){2.5}
\Line(80,175)(50,125)
\Line(250,175)(220,125)
\Line(80,175)(110,125)
\Line(250,175)(280,125)
\ArrowLine(110,125)(50,125)
\ArrowLine(280,125)(220,125)
\Photon(50,125)(50,85){-2}{4}
\Photon(110,125)(110,85){2}{4}
\Photon(220,125)(220,85){-2}{4}
\Photon(280,125)(280,85){2}{4}
\Text(80,185)[]{$\gamma_\lambda\gamma_5$}
\Text(250,185)[]{$\gamma_\lambda\gamma_5$}
\Text(42,105)[]{$p$}
\Text(118,105)[]{$k$}
\Text(212,105)[]{$k$}
\Text(289,105)[]{$p$}
\Text(50,77)[]{$\mu$}
\Text(110,77)[]{$\nu$}
\Text(220,77)[]{$\nu$}
\Text(280,77)[]{$\mu$}
\Text(80,117)[]{$l$}
\Text(250,117)[]{$l$}
\Text(80,50)[]{(i)}
\Text(250,50)[]{(ii)}
\end{picture}
\end{center}

\end{document}